Fig. 1

a) 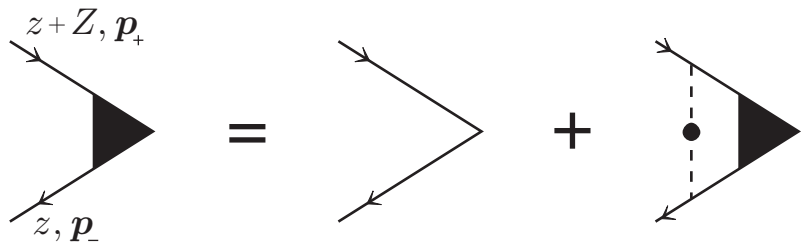

b) 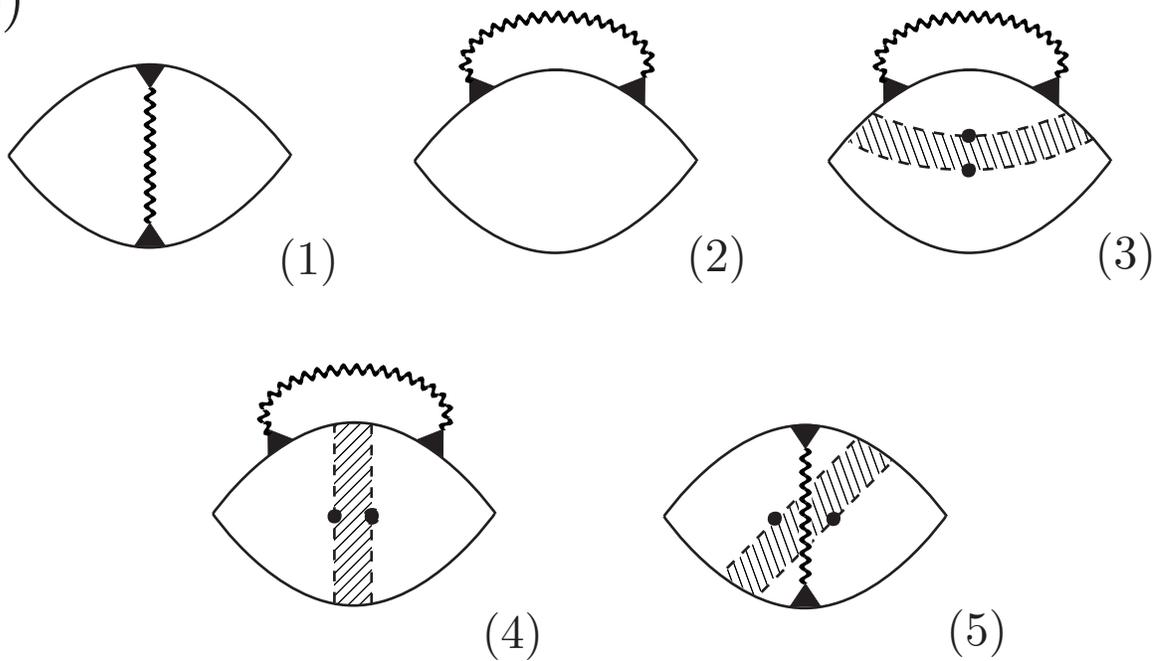

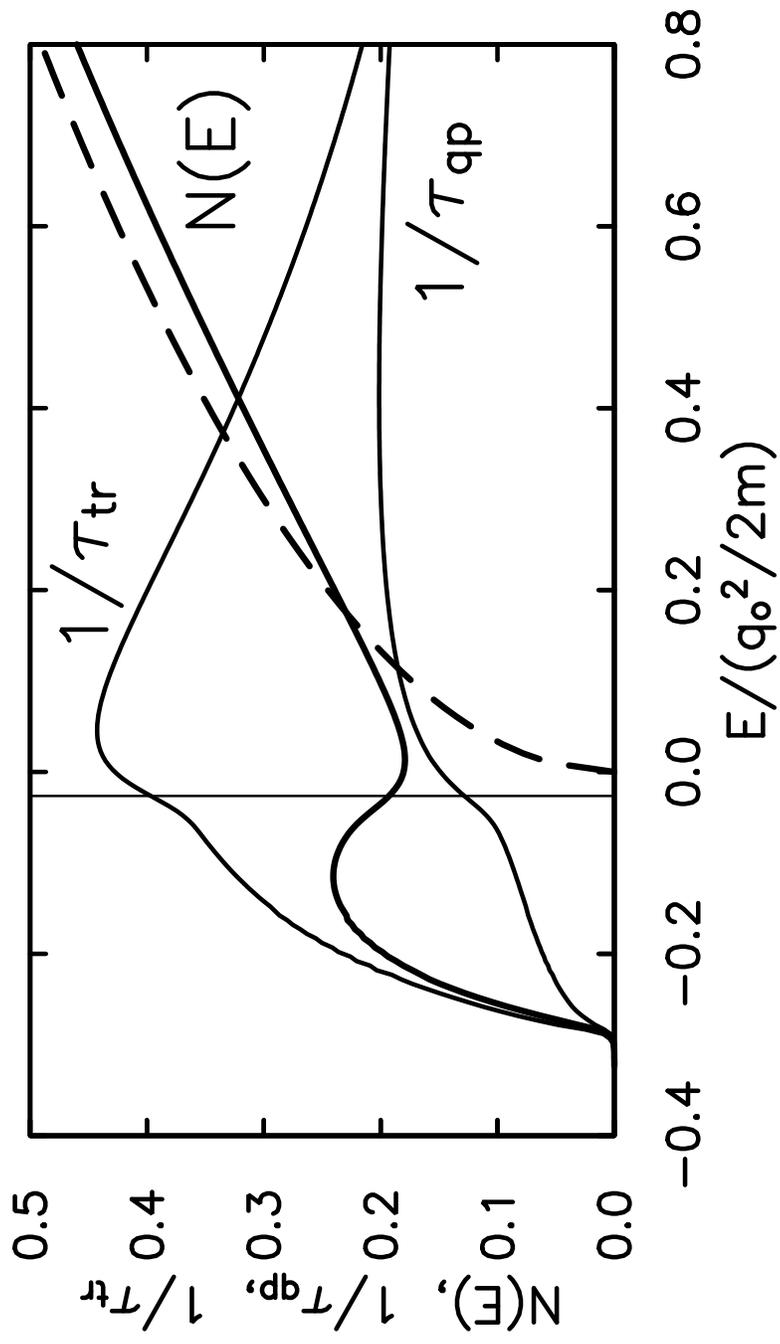

Fig. 2

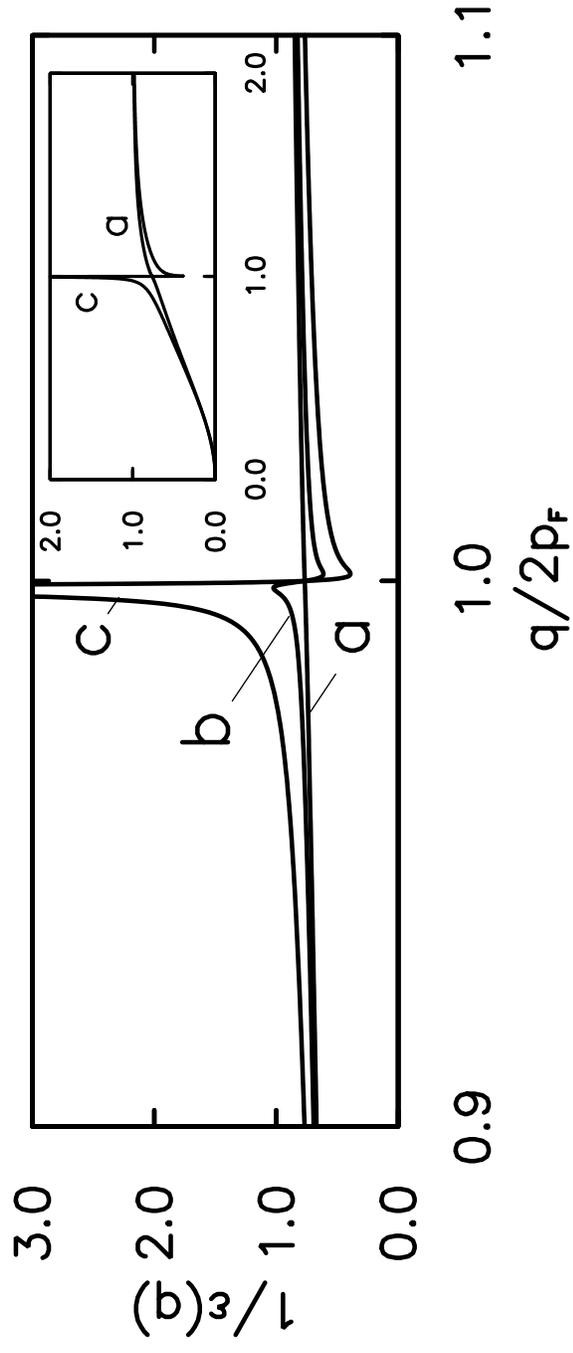

Fig. 3

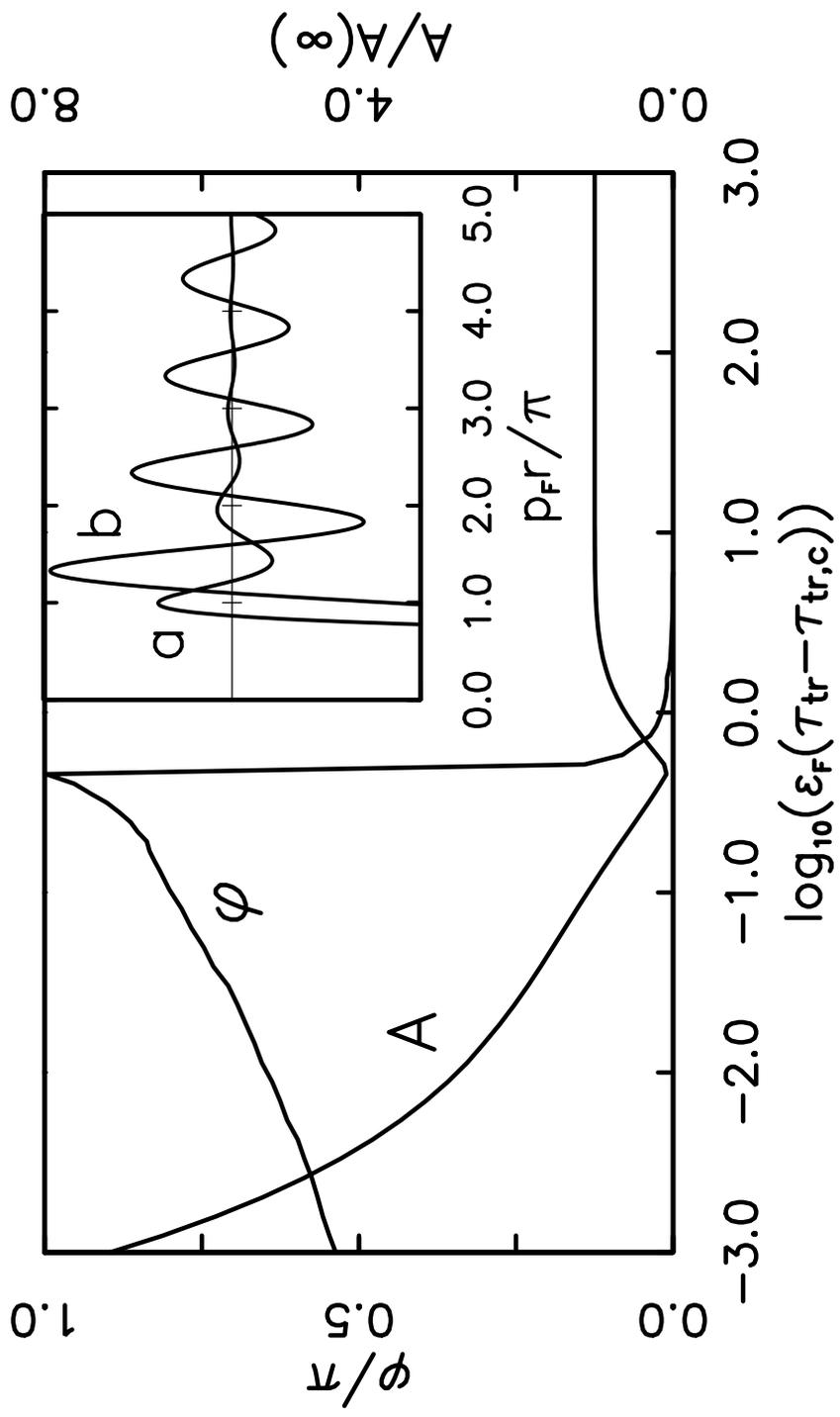

Fig. 4

# Quantum Interference of Coulomb Interaction and Disorder: Phase Shift of Friedel Oscillations and an Instability of the Fermi Sea

Johann Kroha[1], Andrea Huck[2†], and Thilo Kopp[2]

[1] *Laboratory of Atomic and Solid State Physics, Cornell University, Ithaca, NY 14853*

[2] *Institut für Theorie der Kondensierten Materie, Postfach 6980, 76128 Karlsruhe, Germany*

(November 22, 1994)

## Abstract

We investigate the influence of interference between Coulomb interaction and impurity scattering on the static electronic response $\chi(0,q)$ in disordered metals to leading order in the effective Coulomb interaction. When the transport relaxation time $\tau_{tr}$ is much shorter than the quasiparticle life time, we find a $\mathrm{sgn}(2p_F - q)/\sqrt{|2p_F - q|}$ divergence of the polarization function at the Fermi surface ($q = 2p_F$). It causes a phase shift of the Friedel oscillations as well as an enhancement of their amplitude. Our results are consistent with experiments and may be relevant for understanding the stability of the amorphous state of certain alloys against crystallization.

Due to the restriction of electrons to states with momentum less than or equal to the Fermi momentum $p_F$, electronic density correlations exhibit spatial oscillations, generally referred to as Friedel oscillations [1]. It has been suggested [2–4] that the resulting oscillatory contribution $\Phi^{osc}(r)$ to the effective pair potential between the ions in a metal could be important for stabilizing the amorphous state of certain alloys, in that the ions arrange themselves in commensurate positions at the minima of $\Phi^{osc}(r)$. Indeed, experiments on a wide class of quenched noble–polyvalent metal alloys, which exhibit a crystalline to amorphous transformation as a function of noble metal concentration, clearly show a relation between atomic and electronic structure [4,5]: In all those systems the transition occurs, when the Fermi energy $\epsilon_F$ lies in a structure–induced pseudogap of the density of states (DOS). Moreover, in the transition region, the atomic spacing $a$ coincides with the Friedel wave length $\lambda_F = \pi/p_F$ over distances of several $\lambda_F$. However, the atomic positions are shifted by $\lambda_F/4$ compared to the positions of the minima of $\Phi^{osc}(r)$ as calculated for a clean electron sea. Assuming that in the amorphous state the atoms are sitting in the minima of $\Phi^{osc}(r)$, it may be conjectured that in those systems Friedel oscillations acquire a phase shift. The problem of the strength of the Friedel coupling has been a subject of vivid interest. In disordered systems the oscillations of the impurity–averaged potential $\langle \Phi^{osc}(r) \rangle$ are exponentially damped [6] due to a random phase, which arises from the spread of the Fermi wave length over a width given by the elastic mean free path [7,8]. Yet, Jagannathan et.al. [9] showed that higher moments of the distribution of $\Phi^{osc}(r)$ produce a long–range coupling. In this way, the behavior of RKKY spin glass systems [10] at low temperatures $T$ could be explained. In the amorphous systems considered here, however, thermal energies ($T = 77K$) are typically of the same size as the depth of the first minimum of $\Phi^{osc}(r)$ of even a clean electron sea. Thus, the binding of ions by $\Phi^{osc}(r)$ and the existence of a phase shift of Friedel oscillations have remained controversial issues.

In this paper we show that the interplay between Coulomb interaction and impurity scattering can give rise to a phase shift as well as an enhancement of the amplitude of the electron density oscillations in systems where the electronic transport time $\tau_{tr}$ is short compared to the quasiparticle life time $\tau_{qp}$. We give arguments that this condition is satisfied near the crystalline to amorphous transition. The above effect is intimately related to a previously undiscovered instability of a disordered electron sea towards formation of concentrical charge density oscillations in 3 dimensions.

While the response of disordered electron systems has been studied extensively in the hydrodynamic limit [11], Friedel oscillations are governed by the static electronic response $\chi(0, q)$ at momenta $q \simeq 2 p_F$, which is unexplored to a large extent. In a disordered Fermi liquid the interaction vertex is dressed by impurity ladder vertex corrections (Fig. 1(a)) [12]. Due to their diffusion pole structure, any two–particle interaction acquires a strongly retarded, long–range part [13,14,11]. This can drastically alter the screening as shown below. If one neglects any coherent scattering processes, impurity vertex corrections are simply inserted into the vertex of the well–known random phase approximation (RPA). When taken into account on this classical level of approximation, neither substitutional nor structural (amorphous) disorder produces a phase shift of the density oscillations [15]. The reason is that in those diagrams the impurity ladder vertex $\Gamma$ enters in the static limit at finite q, so that the diffusion pole is suppressed. We are thus led to consider vertex corrections where $\Gamma$ is integrated over frequency and momentum, i.e. does give an infrared divergent contribution. There the hydrodynamic transport properties become important, although the response is taken at large external momenta.

The strong isotropic density correlations between the ions of the system [5] have important implications on the electronic transport: The impurity scattering T matrix $t_{\boldsymbol{p},\boldsymbol{p'}}$ has a pronounced peak at a momentum transfer of size $q \equiv |\boldsymbol{p'} - \boldsymbol{p}| = q_o$, where $q_o := 2\pi/a$. This can be seen directly from the ionic structure factor $S(q)$, measured by electron diffraction [5], which is proportional to $t_{\boldsymbol{p},\boldsymbol{p'}}$. The scattering angle $\Theta$ for "on–shell" scattering (i.e. in and outgoing momenta $\boldsymbol{p}$, $\boldsymbol{p'}$ on the Fermi surface) with momentum transfer $q_o$ is given by $\cos \Theta = 1 - 2(q_o/2 p_F)^2$. It is seen that backscattering is enhanced when $q_0 \simeq 2 p_F$, or $a = \lambda_F$, in analogy to Bragg reflection in 1 dimension. The transport rate $\tau_{tr}^{-1}$ is obtained by solving [16] the impurity ladder Bethe–Salpeter equation as $\tau_{tr}^{-1}/\tau_{qp}^{-1} = 1 + \tau_{qp}M$, where M denotes the "current kernel",

$$M = -2\frac{\sum_{\bm{p},\bm{p}'}(\bm{v_p}\hat{\bm{q}})G_{\bm{p}}''(0^-)|t_{\bm{p},\bm{p}'}|^2 G_{\bm{p}'}''^2(0^-)(\bm{v_{p'}}\hat{\bm{q}})}{\sum_{\bm{p}}(\bm{v_p}\hat{\bm{q}})^2 G_{\bm{p}}''^2(0^-)}. \quad (1)$$

$G_{\bm{p}}(0^-)$ and $\bm{v_p}$ are the advanced Green function at the Fermi energy and the group velocity, respectively. A $''$ denotes the imaginary part throughout this paper. We assume $|t_{\bm{p},\bm{p}'}|^2 \propto \delta(|\bm{p}' - \bm{p}| - q_o)$. Because of the peak structure of $G_{\bm{p}}''$ at the Fermi surface, M is governed by on–shell scattering, and in the transition region to the amorphous state ($q_o \simeq 2p_F$) $\tau_{tr}^{-1}$ is enhanced compared to $\tau_{qp}^{-1}$ because of the antisymmetry of the current vertices $(\bm{v_p}\hat{\bm{q}})$. To obtain an estimate of $\tau_{tr}^{-1}/\tau_{qp}^{-1}$ we have evaluated Eq. (1), where the single-particle quantities have been calculated in self–consistent Born approximation. Note that this constitutes a conserving approximation. It reproduces the observed [5] structure–induced pseudogap of the DOS (Fig. 2). Fig. 2 also shows a strong enhancement of $\tau_{tr}^{-1}/\tau_{qp}^{-1}$ when $2p_F \simeq q_o$, or equivalently, when $\epsilon_F$ is near the minimum of the DOS. This will become important below.

We now turn to calculating the polarization function $\Pi(\omega, q)$. All corrections to $\Pi(\omega, q)$ to leading order in the effective Coulomb interaction $\bar{v}_q(z, Z)$ are shown in Fig. 1(b). We note that these diagrams are generated in a conserving approximation from the same class of free energy diagrams that produces, on the single-particle level, the Al'tshuler–Aronov [13] self–energy, which causes a $\sqrt{|E - \epsilon_F|}$ singularity of the DOS in a disordered Fermi liquid, as was first shown by those authors. $\Pi(\omega, q)$ is related to the density response $\chi(\omega, q)$ by $\chi = \Pi/(1 - v_q\Pi)$, with $v_q$, the bare Coulomb interaction. For complex frequencies $z + Z$ and $z$ of in and outgoing lines, $\bar{v}_q(z, Z)$ is given by

$$\bar{v}_q(z, Z) = \frac{v_q}{\epsilon^{RPA}(Z, q)}\Gamma^2(z, Z, q), \qquad v_q = \frac{4\pi e^2}{q^2} \quad (2)$$

where $\epsilon^{RPA}(Z, q) = 1 + 2\pi i\ \sigma\ \text{sgn}Z''/(Z + iq^2 D\ \text{sgn}Z'')$ is the disordered RPA dynamical dielectric function and

$$\Gamma(z, Z, q) = \begin{cases} \frac{i/\tau_{tr}\ \text{sgn}Z''}{Z + iq^2 D\ \text{sgn}Z''} & \text{if } z''(z + Z)'' < 0 \\ 1 & \text{otherwise} \end{cases} \quad (3)$$

is the dressed interaction vertex, $D = 1/3\ v_F^2 \tau_{tr}$ and $\sigma = ne^2\tau_{tr}/m$ denoting the diffusion coefficient and the Drude conductivity, respectively. In Eq. (2) $\epsilon^{RPA}$ and $\Gamma^2$ combine to a single diffusion pole indicating the retarded, long–range character of $\bar{v}_q$. Since the Drude conductivity, expressed in terms of the plasma frequency $\omega_p$ as $\sigma = \omega_p^2 \tau_{tr}$, is typically three orders of magnitude larger than the Fermi energy $\epsilon_F$, the nondivergent terms may be neglected.

It may be shown by Fourier transformation that a phase shift of the Friedel oscillations corresponds to a peak structure of $\chi(0, q)$ at $q = 2p_F$. As seen below, the appearance of such a structure is connected with a divergence of the effective interaction at zero frequency and momentum transfer from the particle to the hole line in $\Pi(0, q)$. Since this divergence is obviously strongest in diagram (1), Fig. 1(b), we only need to consider this diagram, $\Pi^{(1)}$, in order to calculate the leading singular behavior at the Fermi surface ($q = 2p_F$). Note in this context that, while diagrams (1)–(3), Fig. 1(b) cancel each other exactly in the limit $\omega, q \to 0$ [14], this is no longer the case for finite momenta. $\Pi^{(1)}$ reads

$$\Pi^{(1)}(0,q) \equiv -2T^2 \sum_{E,\omega'} \sum_{\boldsymbol{p},\boldsymbol{q}'} G_{\boldsymbol{p}+-}(iE)\ G_{\boldsymbol{p}--}(iE)\ \bar{v}_{q'}(iE, i\omega')\ G_{\boldsymbol{p}++}(iE+i\omega)\ G_{\boldsymbol{p}-+}(iE+i\omega) \quad (4)$$

$$= -2\mathcal{P} \sum_{\boldsymbol{p},\boldsymbol{q}'} \int \frac{d\omega'}{\pi}\ \frac{[f(\omega' + \epsilon_{\boldsymbol{p}+-}) - f(\epsilon_{\boldsymbol{p}+-})]\ \bar{v}''_{q'}(0 + i0, |\omega'| - i0)}{(\epsilon_{\boldsymbol{p}+-} - \epsilon_{\boldsymbol{p}--})(\omega' + \epsilon_{\boldsymbol{p}+-} - \epsilon_{\boldsymbol{p}++})(\omega' + \epsilon_{\boldsymbol{p}+-} - \epsilon_{\boldsymbol{p}-+})} \quad (5)$$

+ nondivergent terms.

Here, $f$ is the Fermi function, $\boldsymbol{p}_{+-}$ stands for $\boldsymbol{p}+\boldsymbol{q}/2-\boldsymbol{q}'/2$ etc., and $\mathcal{P}$ indicates the principal value. The $\omega'$ integral over the singular part of $\bar{v}$ in Eq. (4) extends up to frequencies $\sim (2\pi/3)\tau_{tr}^{-1}$. Near the amorphous transition $\tau_{qp}^{-1}$ is substantially smaller than $\tau_{tr}^{-1}$ (see above). Thus, in evaluating $\Pi^{(1)}$ we have assumed the quasiparticle pole of $G_{\boldsymbol{p}}(z)$ to lie on the real axis for the sake of clarity of the present calculation. Careful inspection of all terms contributing to $\Pi^{(1)}$ shows that only the one written explicitly in Eq. (5) is divergent at the Fermi surface. Due to the strong $q' = 0$ divergence of $\bar{v}''_{q'}$ we can set $q' = 0$ in all nonsingular terms [13]. The momentum integrals can then be performed without further approximations: Applying an integration by parts in the integral over the component of $p$ parallel to $q$ elucidates the fact that $\Pi^{(1)}$ is governed by contributions from the Fermi surface, and we obtain ($T \ll \epsilon_F$)

$$\Pi^{(1)}(0,q) = -\frac{3\sqrt{3}}{16} \frac{1}{(2\epsilon_F \tau_{tr})^{7/2}} \int_{-\epsilon_F}^{\epsilon_F} \frac{d\nu}{x^{3/2}} \frac{1/(4T)}{\cosh^2 \frac{\nu}{2T}} \left[ \int_{-\infty}^{1/4x} \frac{du}{|u|^{5/2}} \left[ F_1(u,x) - F_2(u,x) \right] \right.$$

$$\left. + \int_{1/4x}^{\infty} \frac{du}{|u|^{5/2}} F_1(u,x) \right], \quad (6)$$

$$F_1(u,x) = L(x) - (1 - \frac{u}{x})L(x_1), \qquad F_2(u,x) = \sqrt{1-4xu}\left[L(\tilde{x}) - (1 - \frac{u}{x})L(\tilde{x}_1)\right]$$

where $L(x) \equiv \Pi^{(0)}(0,q) = -2mp_F/(2\pi)^2 \left[1 + (1-x^2)/(2x) \ln|(1+x)/(1-x)|\right]$ is the Lindhard function ($T, \omega = 0$), and $x = x(\nu) = (q/2p_F)/\sqrt{1+\nu/\epsilon_F}$, $x_1 = x - u$, $\tilde{x} = x/\sqrt{1-4xu}$ and $\tilde{x}_1 = (x-u)/\sqrt{1-4xu}$. Each term in the first integral over $u$ has a nonintegrable divergence at $u = 0$, which is compensated only by a cancellation up to $O(u)$ when $F_1$ and $F_2$ are combined. It follows that the $q$ dependence of $\Pi^{(1)}$ is governed by the *second derivative* of $L(x)$, which diverges as $(1-x)^{-1}$. The second $u$ integral in Eq. (6) is nondivergent and may be dropped. To make the $q$ dependence explicit, it is convenient to substitute $u = |x-1|y$. In this way $\Pi^{(1)}$ may be written in the vicinity of $x = 1$ as

$$\Pi^{(1)}(0,q) = -\frac{3\sqrt{3}\, I(1)}{16(2\epsilon_F \tau_{tr})^{7/2}} \int d\nu \frac{1/(4T)}{\cosh^2 \frac{\nu}{2T}} \frac{\text{sgn}(x-1)}{\sqrt{|x-1|}}, \quad (7)$$

where

$$I(x) \equiv \int_{-\infty}^{1/4x|x-1|} \frac{dy}{|y|^{5/2}} \frac{\left[F_1(u,x) - F_2(u,x)\right]\big|_{u=|x-1|y}}{x^{3/2}(x-1)} \quad (8)$$

is a continuous, positive function at $x = 1$. The integral may be evaluated numerically as $I(1) \simeq 11.8(2mp_F/(2\pi)^2)$.

Eq. (7) is the main result of this paper. It is seen that at $T = 0$ the static polarization function $\Pi(0,q) = \Pi^{(0)}(0,q) + \Pi^{(1)}(0,q)$ acquires a powerlaw divergence with a sign change at $q = 2p_F$, the half integer exponent being characteristic for diffusive behavior. Consequently, below a temperature $T_<$ the static dielectric function $\epsilon(q) = 1 - v_q \Pi(0,q)$ becomes less than 1 near $q = 2p_F$, and it follows from a Kramers–Kronig relation that the dynamical stability criterion, $\omega \chi''(\omega - i0, q) \geq 0$, is no longer fulfilled for all frequencies: The disordered electron sea becomes unstable towards formation of charge density oscillations with wave length $\lambda_F = \pi/p_F$. Note however, that a prerequisite for this instability is $\tau_{tr}^{-1} \gg \tau_{qp}^{-1}$, which means in the present case that there are ionic correlations with wave number $q_o = 2p_F$. Hence, the negative $2p_F$ peak in $\chi(0,q)$ is compensated by a peak in the static *ionic* response, which is equal and opposite in sign due to local charge neutrality. Further calculations will be needed to confirm that the ionic and the electronic structures are indeed mutually stabilized in this way.

Our results are summarized in Figs. 3. and 4. The screening charge density $\rho(r)$ is computed as the Fourier transform of $1/\epsilon(q) - 1$ [17]. In $\rho(r)$ the disorder induced oscillations are superimposed on the conventional $\cos(2p_F r - \varphi)$, $\varphi = 0$, Friedel oscillations. As long as $\epsilon(q) \geq 1$, the $2p_F$ powerlaw singularity results in a steplike behavior of $1/\epsilon(q)$ at finite $T$, however with an opposite sign compared to the clean response function (Fig. 3(b)), and hence produces density oscillations $\propto -\cos(2p_F r)$. Thus, there is no significant phase shift $\varphi$ in this region. Rather, the Friedel oscillations are weakened and even become overcompensated by the quantum corrections. In Fig. 4 the latter is seen as the point where the amplitude $A$ vanishes and the phase shift jumps to $\varphi = \pi$. As $\epsilon_F \tau_{tr}$ is further decreased, $1/\epsilon(q)$ develops an increasingly asymmetric peak structure at $2p_F$ (Fig. 3(c)), which leads to a contribution $\propto \sin(2p_F r)$ superimposed on the conventional oscillations. Thus, as the weight of the $2p_F$ peak grows, the amplitude $A$ of the density oscillations increases and $\varphi$ approaches $\pi/2$. Eventually, at a critical transport rate $\tau_{tr,c}^{-1}$, $\epsilon(2p_F)$ becomes 0, and the peak turns into a divergence, i.e. the amplitude of the disorder induced density oscillations diverges, marking the breakdown of perturbation theory. Near that point, the conventional Friedel oscillations are negligible, resulting in a phase shift of $\varphi = \pi/2$. It should be emphasized that, although the divergence of $\Pi(0,q)$, Eq. (7), is changed to a peak when a finite quasiparticle life time $\tau_{qp}$ is included, the divergence of $1/\epsilon(q)$ persists as long as $\tau_{tr}^{-1} \gg \tau_{qp}^{-1}$. From the above discussion it is now clear that $\varphi \simeq \pi/2$ is intimately related to a strong enhancement of the amplitude of the oscillations. This explains the curious experimental finding [4,5] mentioned in the introduction that the transition to the amorphous state occurs at a shift of the atomic positions by $\lambda_F/4$, corresponding to $\varphi = \pi/2$, because at that same point the effective pair potential $\Phi^{osc}(r)$ becomes particularly strong and dominates the crystal potential and thermal energies.

In conclusion, we have calculated the static electronic response of a disordered metal, including quantum interference between Coulomb interaction and impurity scattering to leading order in the effective Coulomb interaction. In a situation where the transport relaxation rate is much greater than the quasiparticle relaxation rate, the quantum corrections display a previously unknown powerlaw divergence at $q = 2p_F$. This structure leads to a phase shift as well as to an enhancement of the amplitude of the Friedel oscillations. Hence, this effect seems to explain experiments [4,5] on certain strongly disordered alloys, and, in particular, may act stabilizing on the amorphous state of these materials. A detailed exposition of the calculations presented here, including finite quasiparticle life time will be presented elsewhere.

We are grateful to P. Wölfle, A. A. Aronov, N. W. Ashcroft, and P. Häussler for valuable discussions and comments. This work was supported in part by SFB 195 (T.K.) and by the Alexander von Humboldt Foundation (J.K.).

† Present address: I. Inst. für Theor. Physik, Universität Hamburg, Jungiusstraße 9, 20355 Hamburg, Germany

FIGURES

FIG. 1. (a) Impurity dressing of the interaction vertex. (b) All corrections to the polarization function $\Pi(\omega, q)$ in leading order in the effective Coulomb interaction $\bar{v}_q$ are shown. $\bar{v}_q$ (Eq. (2)) is denoted by a bold wavy line with solid triangles at the ends. The shaded blocks represent diffusion vertices. In diagrams (2)–(5) a corresponding contribution is obtained by interchanging the particle and the hole line.

FIG. 2. $\tau_{qp}^{-1}$, $\tau_{tr}^{-1}$ in units of $q_o^2/2m$, and the DOS $N(E)$ (arb. units) as a function of energy. A quadratic band $E = p^2/2m$ is assumed for the clean electron sea, and parameters are chosen such that $\tau_{qp}^{-1} \simeq 0.12\epsilon_F$ at the Fermi surface. $----$: clean DOS $N^{(0)}(E)$. The vertical line indicates the position of the Fermi level for $2p_F = q_o$.

FIG. 3. $1/\epsilon(q)$ is shown for a clean (a) and a strongly disordered ($\epsilon_F \tau_{tr} = 1.8$ (b) and 1.23 (c)) electron liquid as calculated from Eq. (7) near $q/2p_F = 1$. All numerical evaluations are done with an electron gas density parameter [17] of $r_s = 4$ and at $T/\epsilon_F = 0.001$.

FIG. 4. Phase shift $\varphi$ and amplitude $A$ of the 1st maximum of $\rho(r)$ as a function of $\epsilon_F(\tau_{tr} - \tau_{tr,c})$; $\epsilon_F \tau_{tr,c} = 1.217$. The inset shows $\rho(r)$ (arb. units) as a function of distance $r$ from an impurity ion (a) in the clean and (b) in the strongly disordered case ($\epsilon_F \tau_{tr} = 1.23$). The phase shift and the enhancement of the amplitude are clearly seen.